\documentclass[preprint,showpacs,showkeys,preprintnumbers,amsmath,amssymb]{revtex4}

\usepackage{graphicx}
\usepackage{dcolumn}
\usepackage{bm}

\begin{document}

\title{Surface Effects on Oxide Heterostructures}

\author{U.~Schwingenschl\"ogl and C.~Schuster}
\affiliation{Institut f\"ur Physik, Universit\"at Augsburg, 86135 Augsburg,
Germany}

\date{\today}

\begin{abstract}
We report on surface effects on the electronic properties of interfaces
in epitaxial LaAlO$_3$/SrTiO$_3$ heterostructures. Our results are based
on first-principles electronic structure calculations for well-relaxed
multilayer configurations, terminated by an ultrathin LaAlO$_3$ surface
layer. On varying the thickness of this layer, we find that the interface
conduction states are subject to almost rigid band shifts due to a modified
Fermi energy. Confirming experimental data, the electronic properties
of heterointerfaces therefore can be tuned systematically by alterating
the surface-interface distance. We expect that this mechanism is very
general and applies to most oxide heterostructures.
\end{abstract}


\maketitle

\begin{figure}
\includegraphics[width=0.25\textwidth]{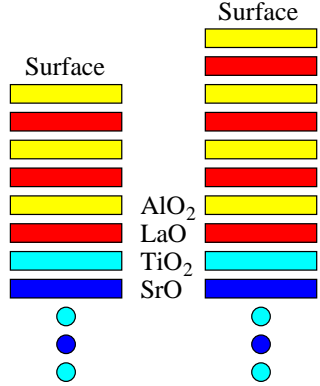}
\caption{(Color online) LaAlO$_3$/SrTiO$_3$ (001) heterostructure
terminated by an (AlO$_2$)$^-$ surface, where the bulk is formed by
alternating (TiO$_2$)$^0$ and (SrO)$^0$ layers. The band structure calculations
presented in this paper refer to surface layers comprising 3 and 4
LaAlO$_3$ unit cells.}
\label{fig1}
\end{figure}

Heterostructures based on II-IV semiconductors have attracted great
attention in recent years due to a large number of possible
applications for electronic devices \cite{kroemer01}. Special interest
has focussed on perovskite heterostructures from transition metal oxides,
because of both the variability of the perovskite structure and
the correlated electron properties associated with transition metals
\cite{imada98}. Pulsed laser deposition techniques and molecular beam
epitaxy nowadays allow to grow layered structures with a precision of
one unit cell. It consequently is possible to create atomically
sharp interfaces between perovskite oxides with strongly different
electronic features, opening a huge field of new cooperative phenomena.
Charge redistribution plays the key role for the electronic structure
at the interface since electrons flow from one layer to another,
driven by different electrochemical potentials in the component materials
\cite{okamoto04}. The physical properties of such a multilayer structure
hence may not be present in either of its constituents. For instance,
a conducting quasi two-dimensional electron gas is found to be formed
between the Mott insulator LaTiO$_3$ and the band insulator SrTiO$_3$
\cite{ohtomo02}.

A contact between two common band insulators with the large band gaps
5.6\,eV and 3.2\,eV is realized in the LaAlO$_3$/SrTiO$_3$ heterointerface.
The structure consists of alternating (SrO)$^0$, (TiO$_2$)$^0$, (LaO)$^+$,
and (AlO$_2$)$^-$ layers, see the schematic representation in figure
\ref{fig1}. At the electron-doped (TiO$_2$)$^0$/(LaO)$^+$ contact a quasi
two-dimensional electron gas with very high carrier density
\cite{ohtomo04,ohtomo06} is formed. In contrast to the mixed valence system
LaTiO$_3$/SrTiO$_3$, the formation of the electron gas here is considered
to result from a polarity discontinuity. Because of different formal
valences of the metal ions Ti$^{4+}$ and La$^{3+}$, uncompensated charge
is left at the interface, giving rise to the metallic layer. Oxygen
vacancies introduced in the SrTiO$_3$ layer during the deposition
process likewise are a source of doping \cite{kalabukhov07}. Thiel
{\it et al.} \cite{thiel06} have studied the transport properties of
the LaAlO$_3$/SrTiO$_3$ interface using the electric field-effect.
Measurements on samples without indications of oxygen defects show
that the LaAlO$_3$ surface layer must reach a critical thickness of
4 unit cells for the interface to be conducting.

\begin{figure*}
\includegraphics[width=0.4\textwidth,clip]{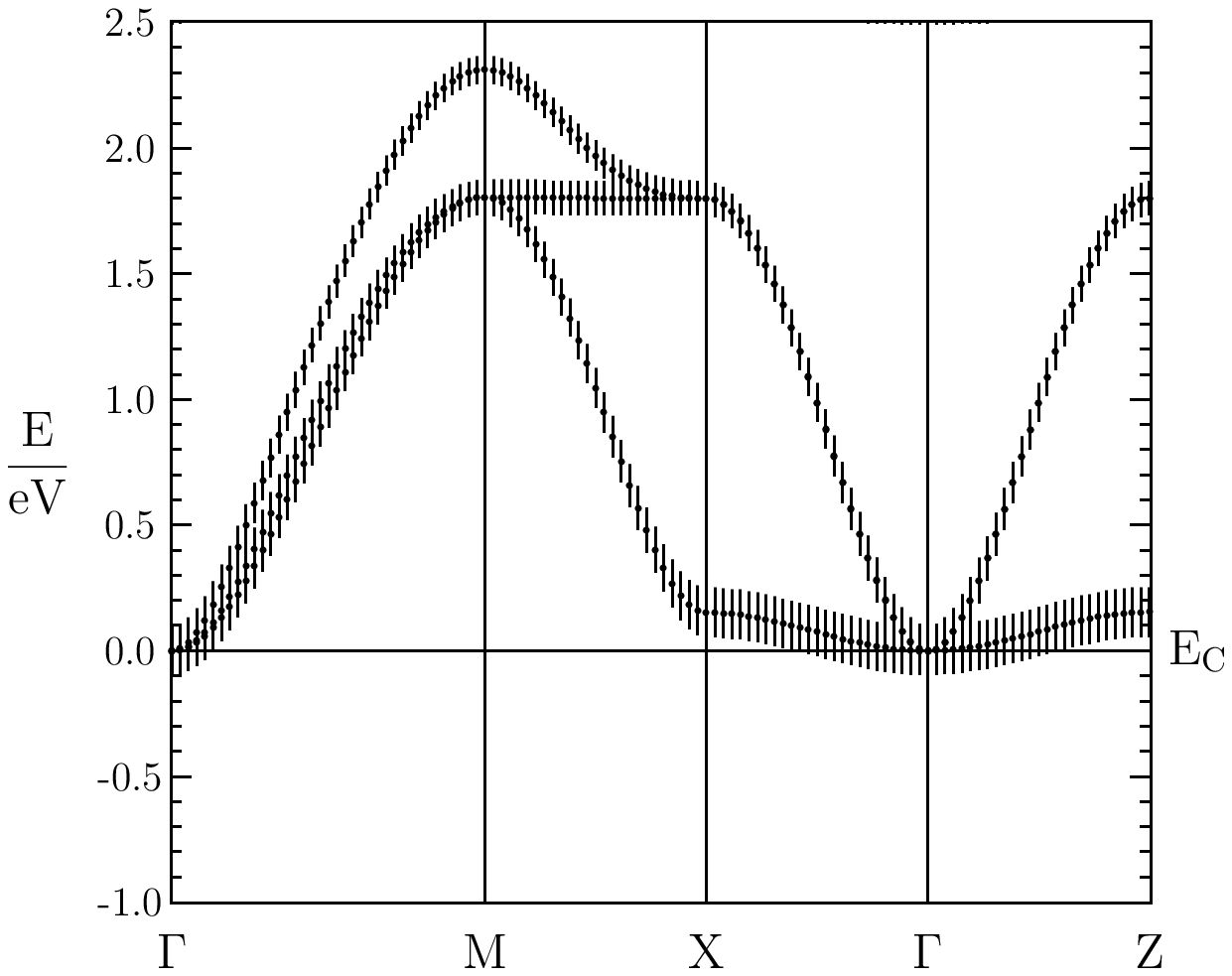}\hspace{1cm}
\includegraphics[width=0.4\textwidth,clip]{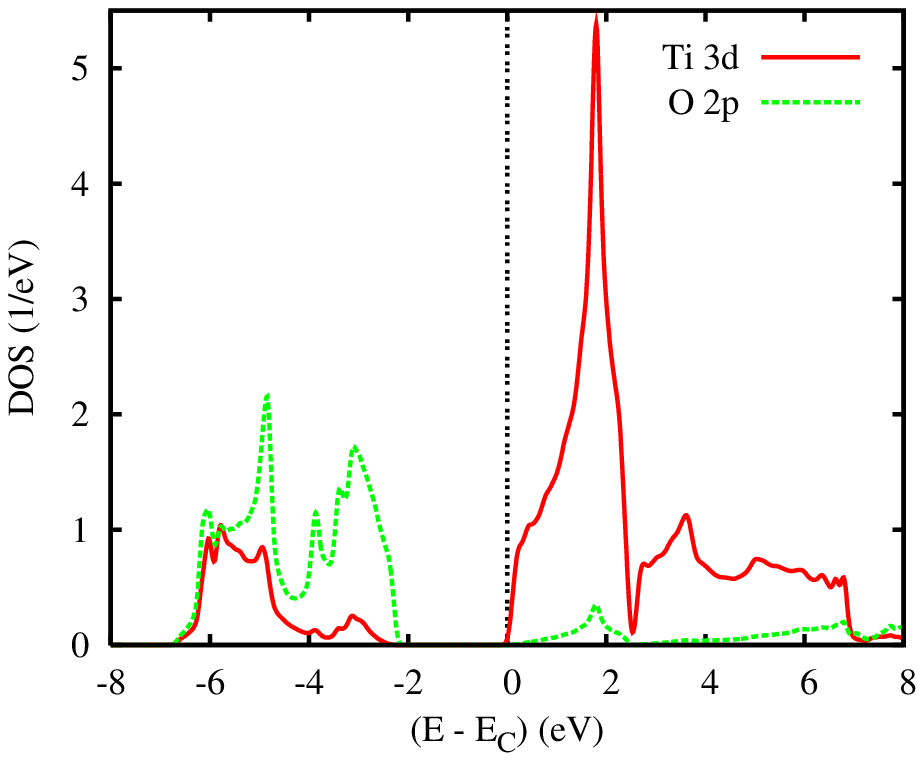}
\caption{(Color online) Band structure and partial Ti $3d$ and O $2p$
DOS (per atomic site) for bulk SrTiO$_3$. The weighted bands highlight
contributions of the Ti $3d$ $t_{2g}$ states.}
\label{fig2}
\end{figure*}

From the theoretical point of view, charge compensation at
LaAlO$_3$/SrTiO$_3$ interfaces has been investigated by Park
{\it et al.} \cite{park06}. Electron doping of Ti atoms in combination
with the Jahn-Teller effect leads to metallicity in their band
structure calculations, based on the local density approximation (LDA).
In addition, applying the LDA+U scheme, Pentcheva and Pickett \cite{pentcheva06}
have found ferromagnetically aligned spins due to occupied $d_{xy}$ orbitals in
a checkerboard of Ti$^{3+}$ sites. Beyond structural relaxation effects at
the heterointerface, serious modifications of the electronic structure
are caused by surfaces \cite{us07a,us07b}. For the LaAlO$_3$ (001)
surface, atomic reconfiguration due to the polarity discontinuity
at the vacuum interface is accompanied by a remarkable redistribution of
the surface electron density, see Lanier {\it et al.} \cite{lanier07}
and the references therein. In spite of the relevance of these results
for heterointerfaces near surfaces, first principles band structure
calculations accounting for the vacuum interface are missing so far.
Our present work therefore addresses LaAlO$_3$/SrTiO$_3$ multilayer
configurations terminated by an (AlO$_2$)$^-$ surface. We discuss the
interplay between the electronic states at the heterointerface and
their distance from the surface. Modifying this distance turns out to
be a most useful instrument for systematically tuning the transport
properties.

The band structure results discussed in the following rely on the
scalar-relativistic augmented spherical wave (ASW) approach \cite{eyert00}.
We apply a recently implemented code which accounts for the non-spherical
contributions to the charge density inside the atomic spheres \cite{eyert07}.
The ASW method is advantageous for dealing with complex structures
comprising a large number of atoms \cite{us04,us06,us07}. We model
heterointerfaces by long tetragonal supercells based on the parent
perovskite structure. The principal axis of these cells gives rise the
crystallographical $c$-axis, which runs perpendicular to the surface.
To be specific, a supercell comprises a layer of 4 SrTiO$_3$ unit cells
sandwiched between layers of 3--4 LaAlO$_3$ unit cells, compare figure
\ref{fig1}. The LaAlO$_3$ layers themselves are terminated by a vacuum
layer of at least 12\,\AA\ thickness, allowing us to apply periodic boundary conditions. As superstructures
perpendicular to the interface have not been reported, they are not taken
into account. Moreover, we use the symmetric crystal structure of
high temperature cubic SrTiO$_3$ for the whole supercell.
Lattice mismatch between SrTiO$_3$ and LaAlO$_3$ layers has to be
avoided via a common lattice constant. Since LaAlO$_3$ is grown on a
SrTiO$_3$ substrate in the experiment, it is reasonable to choose the
measured value for cubic SrTiO$_3$, which amounts to 3.91\,\AA\
\cite{lb76}. In the LaAlO$_3$ region, this approximation artificially
elongates the lattice constant by (less than) 2.5\% with respect to the
experimental value of 3.81\,\AA\ \cite{lb76}. Our lattice setup is
supported by experimental data as well as theoretical structure optimization.

\begin{table}
\begin{tabular}{l|c|c}
&interface&\hspace*{0.5cm}surface\hspace*{0.5cm}\\\hline
Al--O & 1.78\,\AA & 1.92\,\AA \\
La--O$_{\rm Al}$ & 2.76\,\AA & 2.75\,\AA\\\hline
La--O$_{\rm Ti}$ & 2.72\,\AA &\\
Ti--O$_{\rm La}$ & 2.06\,\AA &\\\hline
Ti--O$_{\rm Sr}$ & 1.88\,\AA &\\
Sr--O & 2.74\,\AA, 2.83\,\AA &\\
\end{tabular}
\caption{Metal--O bond lengths at the surface and the interface. The
names of the O sites refer to the layer they are located in. For the Sr--O
bond length the short/long bond points off/towards the interface.
The structure optimization has started from bond lengths of
1.95\,\AA\ (Ti/Al--O) and 2.76\,\AA\ (Sr/La--O), while the intraplanar
metal--O distances keep their original values.\label{tab1}}
\end{table}
Structural relaxation at both the LaAlO$_3$/SrTiO$_3$ interface and
the surface is taken into account via a minimization of the
atomic forces. For this purpose we use the Wien2k program package \cite{wien2k}. 
This is a famous full-potential linearized augmented plane wave code having shown
great capability in the optimization of surfaces as well as interfaces
\cite{us07c,us07d}. The surface relaxation starts out from
a supercell consisting of two LaAlO$_3$ unit cells along the $c$-axis,
terminated by a vacuum layer. As compared to the bulk atomic positions,
the surface (AlO$_2$)$^-$ layer developes a distinct buckling, because
the O atoms are shifted to the vacuum by 0.02\,\AA\ and the Al
atoms move 0.05\,\AA\ in the opposite direction. In the adjacent
(LaO)$^+$ layer, the La atoms follow the surface O sites by 0.03\,\AA,
while the O atoms retain their bulk positions. Turning to the
relaxation effects at the heterointerface, we use a supercell of touching
LaAlO$_3$ and SrTiO$_3$ layers, each covering 4 perovskite unit
cells in $c$-direction. In the (LaO)$^+$ layer at the interface, only the O
atoms move 0.07\,\AA\ off the contact, pushing neighbouring Al atoms
0.02\,\AA\ in the same direction. In contrast, structural modifications
are more pronounced in the (TiO$_2$)$^0$ interface layer: Whereas the
Ti atoms are shifted off the contact by 0.04\,\AA, the O atoms approach
it by 0.06\,\AA. Finally, the neighbouring (SrO)$^0$ layer reveals shifts
of the Sr and O atoms by 0.03\,\AA\ off and towards the contact, respectively.
These findings as well as the optimized bond lengths summarized in
Table \ref{tab1} agree well with the structural data reported by Park
{\it et al.} \cite{park06}.

\begin{figure*}
(a) {\bf Thickness of the LaAlO$_3$ surface layer: 3 unit cells}\\[0.2cm]
\includegraphics[width=0.4\textwidth,clip]{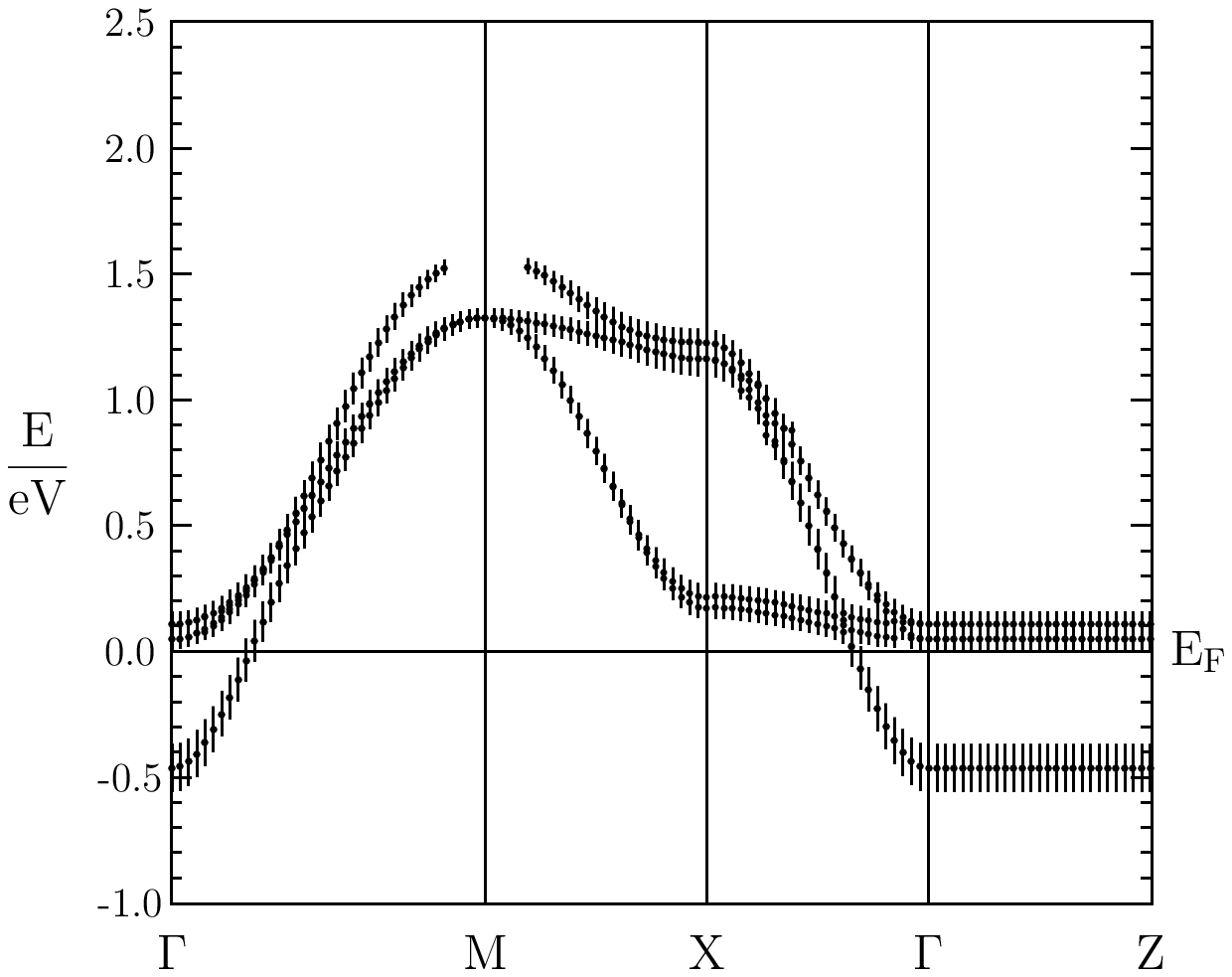}\hspace{1cm}
\includegraphics[width=0.4\textwidth,clip]{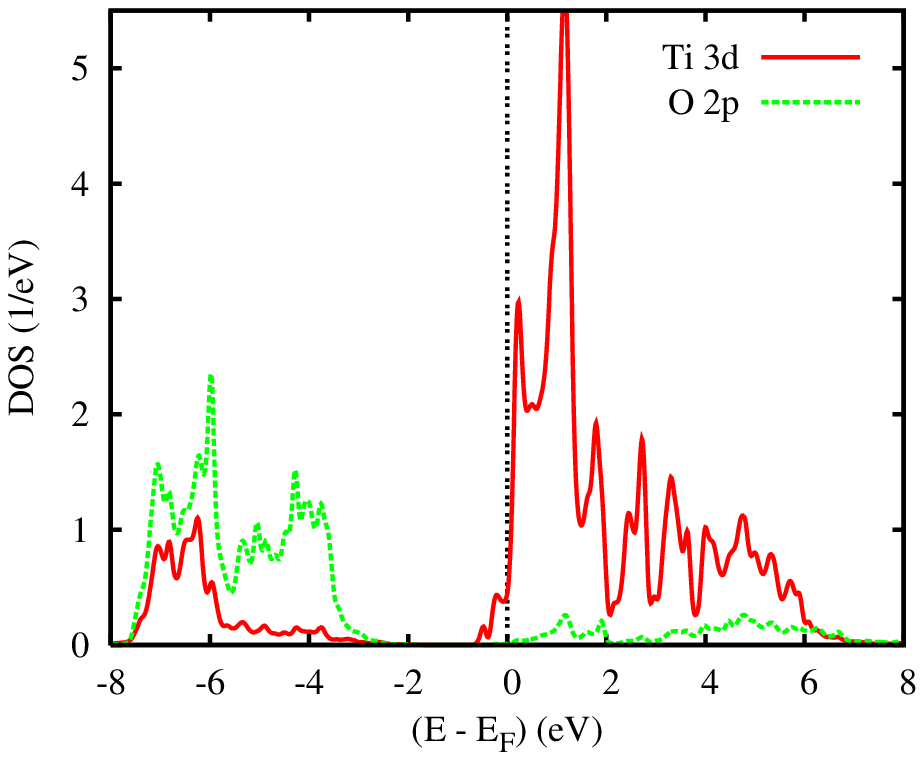}\\[0.2cm]
(b) {\bf Thickness of the LaAlO$_3$ surface layer: 4 unit cells}\\[0.2cm]
\includegraphics[width=0.4\textwidth,clip]{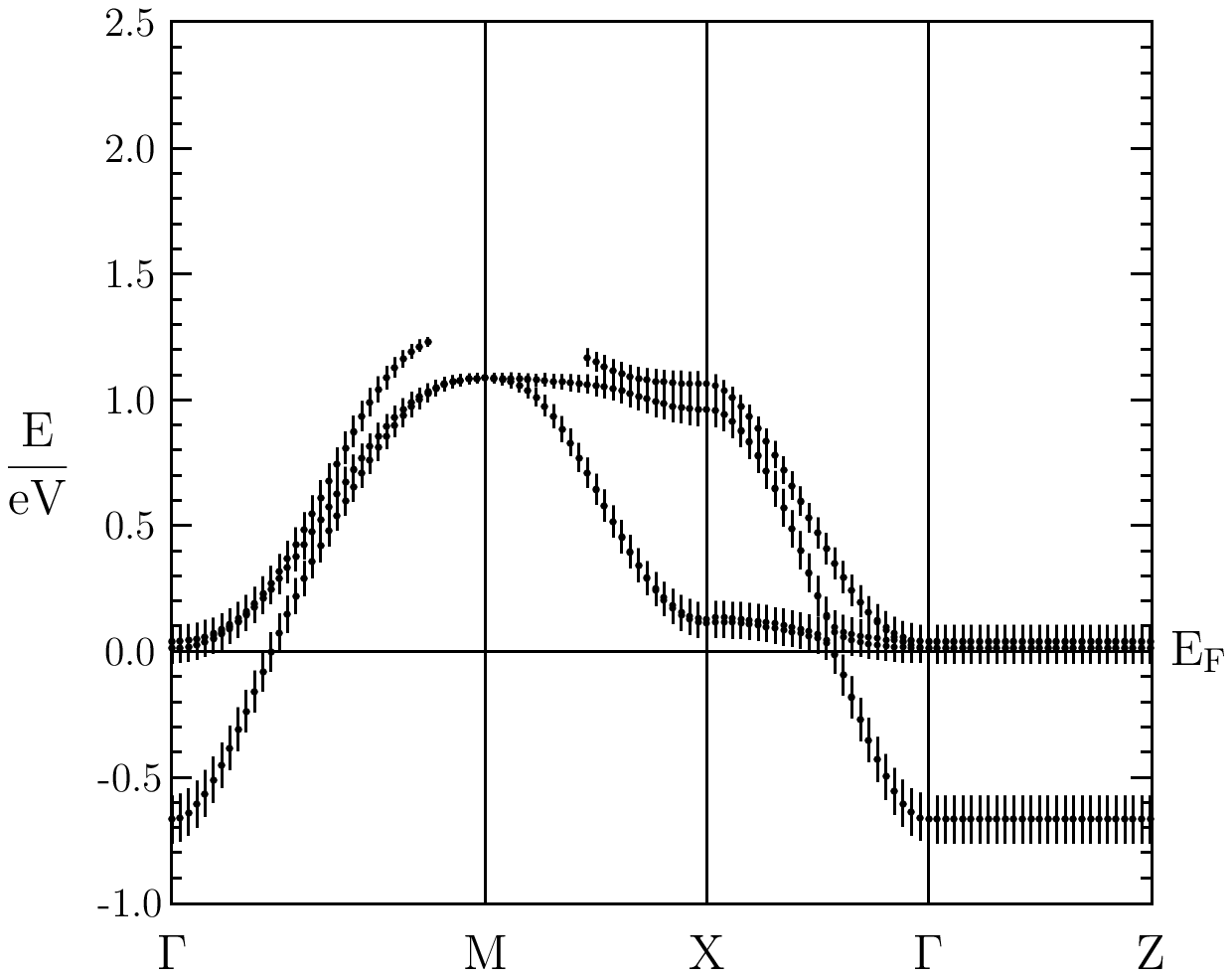}\hspace{1cm}
\includegraphics[width=0.4\textwidth,clip]{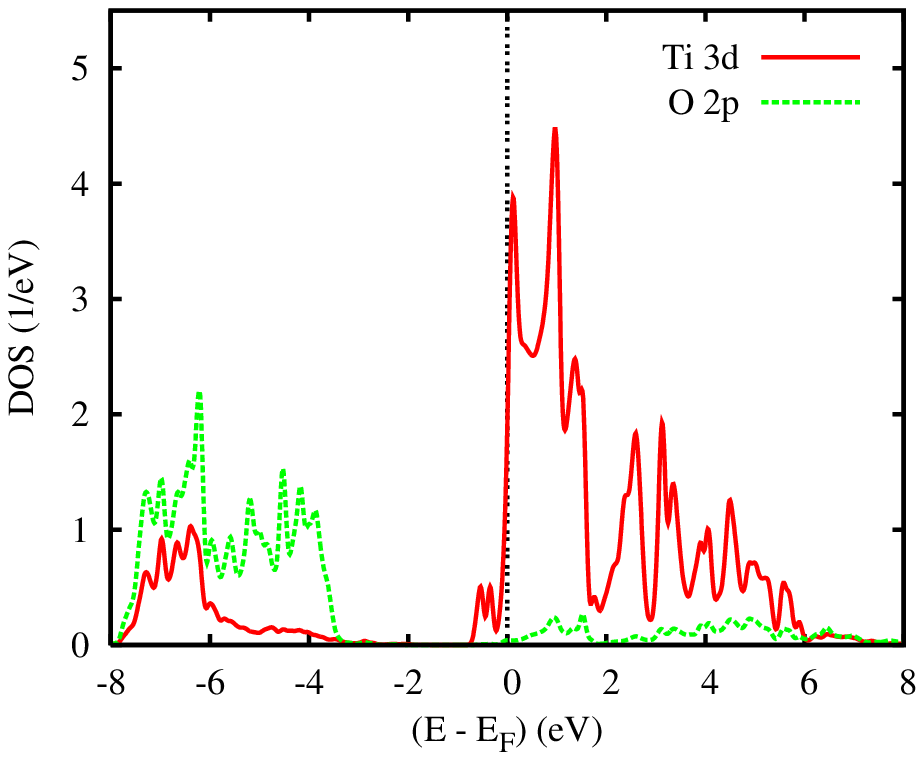}
\caption{(Color online) Band structure and partial Ti $3d$ and O $2p$
DOS (per atomic site) for the interface (TiO$_2$)$^0$ layer, which
is separated by (a) 3 and (b) 4 LaAlO$_3$ unit cells from the surface,
respectively. The weighted bands refer to the simple cubic
Brillouin zone of the parent perovskite structure, showing only
states with Ti $3d$ contributions beyond a certain threshold.}
\label{fig3}
\end{figure*}

For the structure optimization we use the generalized gradient approximation
with a mixed linear-augmented-plane-wave and augmented-plane-wave plus
local-orbitals basis. In the surface and contact
calculation the charge density is represented via about 4{,}000 and
11{,}000 plane waves, respectively, and the {\bf k}-space grid has
10 and 15 {\bf k}-points in the irreducible wedge of the Brillouin zone.
In addition, the Perdew-Burke-Ernzernhof parametrization of the
exchange-correlation potential is applied. On the contrary, the ASW
supercell calculation uses the local density approximation within the
Vosko-Wilk-Nusair scheme.
Aiming at a correct representation of the crystal potential in large voids
of the supercell and near the surface, the physical atomic spheres are
complemented by additional augmentation spheres at carefully selected
interstitial sites. In the case that the surface layer comprises
3 and 4 LaAlO$_3$ unit cells, respectively, it is sufficient to
dispose 153 and 173 additional spheres in order to keep the linear
overlap of the physical spheres below 17\% and of any pair of spheres
below 22\%. Along with 58 and 68 physical spheres, the supercells
entering the ASW calculation thus comprise 211 and 241 augmentation
spheres in total. The basis set taken into account in the secular matrix
in each case consists of La $6s$, $6p$, $5d$, $4f$, Sr $5s$, $5p$,
$4d$, Ti $4s$, $4p$, $3d$, and O $2s$, $2p$ orbitals, as well as
states of the additional augmentation spheres. During the course of
the band calculation the Brillouin zone is sampled with an increasing
number of up to 110 {\bf k}-points in the irreducible wedge, which
ensures convergence of the findings with respect to the fineness of the
{\bf k}-space grid.

To start the discussion of the electronic structure results, we first analyze
bulk SrTiO$_3$ for comparison. The band structure is depicted in figure
\ref{fig2} along selected high symmetry lines in the first Brillouin
zone of the simple cubic lattice, where the symmetry points are defined
by standard reciprocal vectors: $\Gamma=(0,0,0)$, $M=(\frac{1}{2},\frac{1}{2},0)$,
$X=(0,\frac{1}{2},0)$, and $Z=(0,0,\frac{1}{2})$. Contributions of
Ti $3d$ $t_{2g}$ states are represented by the length of the bars added to every
band and {\bf k}-point. Since the system is insulating, the conduction
band is empty, with the minimum at the $\Gamma$-point. One of the
bands giving rise to the conduction band minimum has hardly any dispersion
along the lines $\Gamma$--$X$ and $\Gamma$--$Z$, which are degenerate by
symmetry, thus along the cubic lattice vectors. The remaining states
are characterized by strong dispersion, reflected by the total band width
of about 2.3\,eV. Corresponding densities of states (DOS) are given
on the right hand side of figure \ref{fig2}. In the energy range shown,
contributions of Sr states almost vanish. As to be expected from
a simple molecular orbital picture, the occupied and unoccupied
group of bands is dominated by O $2p$ and Ti $3d$ contributions, respectively,
since orbital overlap leads to bonding and antibonding molecular states.
Admixtures in the energy range dominated by the respective other
states are due to TiO-hybridization, and thus are observed mainly
for $\sigma$-type overlap. These findings agree well with
previous band structure data, see \cite{king-smith94}, for example.

Turning to the LaAlO$_3$/SrTiO$_3$ interface, band structures and DOS
curves are given in figure \ref{fig3} in the same representation as
used for bulk SrTiO$_3$, see figure \ref{fig2}. In particular, the
weighted bands refer to the simple cubic lattice of the parent
perovskite structure. For clarity, we show only such states with
sufficient Ti $3d$ contributions from the electron-doped (TiO$_2$)$^0$
layer right at the interface. Figure \ref{fig3} compares the data
obtained for different surface-interface distances: The thickness of
the LaAlO$_3$ surface layer amounts to (a) 3 and (b) 4 perovskite unit
cells. As concerns the gross features of the band structure as well as
the DOS, the contact Ti $3d$ states largely resemble the bulk SrTiO$_3$
findings. However, the group of conduction bands now is filled
partially, which leaves the interface in a conducting state. To be more
specific, dispersion disappears completely along the line $\Gamma$--Z
due to the spacial restriction induced by the surface. Moreover,
since the single occupied Ti $3d$ band shows strong dispersion within
the $ab$-plane, it belongs to nearly ideal two-dimensional states.
They are visible in the DOS just below the Fermi energy.

Our results confirm the observation of Thiel {\it et al.} \cite{thiel06}
that a growing thickness of the LaAlO$_3$ surface layer enhances the
conductivity of the heterointerface. Because surface electronic states
leak out into the vacuum area, the charge carrier density is reduced
in the LaAlO$_3$ surface layer. As a consequence, less charge is available
for doping the interface. When the distance between interface and
surface increases, the charge depletion declines because of electronic
screening. This trend is obvious in figure \ref{fig3}. For the thinner
LaAlO$_3$ surface layer the two-dimensional band reaches down to
$-0.5$\,eV, whereas the minimum is found at $-0.7$\,eV in the case of
the thicker layer. The occupation therefore increases systematically
with the number of LaAlO$_3$ cells separating the vacuum from
the heterointerface.

As concerns the qualitative dependence on the thickness of the LaAlO$_3$
layer, theory and experiment agree excellent, while a critical number
of 4 unit cells for conductivity is not reproduced by the calculation.
Instead, conductivity is likewise achieved in the case of 3 unit cells,
and even for thinner surface layers, as checked by additional calculations.
This discrepancy probably is connected to the underestimation of the
band gap in pure SrTiO$_3$, tracing back to the local density
approximation. In figure \ref{fig2}, the gap amounts to some $2.1$\,eV,
which is significantly less than the experimental value 3.2\,eV.
Similar shortcomings are known for TiO$_2$ \cite{hardman94} and LaAlO$_3$
\cite{gemming06}. As a consequence of the underestimated band gap, the
occupation of the two-dimensional band at the heterointerface is
overestimated. The transition from insulating to conducting behaviour
therefore cannot not be caught quantitatively. However, these shortcomings
do not affect our previous conclusions about surface effects on
heterointerfaces. In fact, the dependence of the interface charging
on the surface layer thickness is hardly affected by the structural
relaxation, since largely the same behaviour is found for non-relaxed
interfaces. This strongly supports an interpretation in terms of
charge depletion due to the proximity to the surface. In addition,
electronic correlations would not be expected to be critical for the
mechanism under consideration. Our results thus do not depend on
the specific compounds giving rise to the interface.

In conclusion, we have presented first-principles band structure results
for LaAlO$_3$/SrTiO$_3$ heterointerfaces in the vicinity of a surface. As
a consequence of electron-doping due to charge transfer from the
LaAlO$_3$ surface layer, we find the interface (TiO$_2$)$^0$ layer to
be conducting. As the dispersion of the conduction band is strong
and reveals an ideal two-dimensional character,
a quasi two-dimensional gas of very mobile electrons is formed.
In the case of an ultrathin LaAlO$_3$ layer, surface effects are of
special importance for the electronic structure at the heterointerface.
Since electronic states leak out into the vacuum, charge is distributed
close to a surface. For the heterointerface, this implies a rigid
shift of the electronic bands because of a modified Fermi level.
In other words, charge transfer towards a surface counteracts an
intrinsic interface doping. Since this effect declines with the
surface-interface distance, the charge carrier density can be
adjusted via the thickness of the surface layer.

\subsection*{Acknowledgement}
We gratefully acknowledge discussions with U.\ Eckern,
V.\ Eyert, and T.\ Kopp. The work was supported by the Deutsche
Forschungsgemeinschaft (SFB 484).


\begin{thebibliography}{10}

\bibitem{kroemer01}
H.\ Kroemer, Rev.\ Mod.\ Phys.\ {\bf 73}, 783 (2001).

\bibitem{imada98}
M.\ Imada, A.\ Fujimori, and Y.\ Tokura,
Rev.\ Mod.\ Phys.\ {\bf 70}, 1039 (1998).

\bibitem{okamoto04}
S.\ Okamoto and A.J.\ Millis, Nature {\bf 428}, 630 (2004).

\bibitem{ohtomo02}
A.\ Ohtomo, D.A.\ Muller, J.L.\ Grazul, and H.Y.\ Hwang,
Nature {\bf 419}, 378 (2002).

\bibitem{ohtomo04}
A.\ Ohtomo and H.Y.\ Hwang, Nature {\bf 427}, 423 (2004).

\bibitem{ohtomo06}
A.\ Ohtomo and H.Y.\ Hwang, Nature {\bf 441}, 120 (2006).

\bibitem{kalabukhov07}
A.S.\ Kalabukhov, R.\ Gunnarsson, J.\ B\"orjesson, E.\ Olsson,
T.\ Claeson, and D.\ Winkler, Phys.\ Rev.\ B {\bf 75}, 121404(R) (2007).

\bibitem{thiel06}
S.\ Thiel, G.\ Hammerl, A.\ Schmehl, C.W.\ Schneider, and J.\ Mannhart,
Science {\bf 313}, 1942 (2006).

\bibitem{park06}
M.S.\ Park, S.H.\ Rhim, and A.J.\ Freeman,
Phys.\ Rev.\ B {\bf 74}, 205416 (2006).

\bibitem{pentcheva06}
R.\ Pentcheva and W.E.\ Pickett,
Phys.\ Rev.\ B {\bf 74}, 035112 (2006).

\bibitem{us07a}
U.\ Schwingenschl\"ogl and C.\ Schuster,
Chem.\ Phys.\ Lett.\ {\bf 435}, 100 (2007).

\bibitem{us07b}
U.\ Schwingenschl\"ogl and C.\ Schuster,
Chem.\ Phys.\ Lett.\ {\bf 439}, 143 (2007).

\bibitem{lanier07}
C.H.\ Lanier, J.M.\ Rondinelli, B.\ Deng, R.\ Kilaas, K.R.\ Poeppelmeier,
and L.D.\ Marks, Phys.\ Rev.\ Lett.\ {\bf 98}, 086102 (2007).

\bibitem{eyert00}
V.\ Eyert, Int.\ J.\ Quantum Chem.\ {\bf 77}, 1007 (2000).

\bibitem{eyert07}
V.\ Eyert, The Augmented Spherical Wave Method -- A Comprehensive
Treatment, Lecture Notes in Physics (Springer, Heidelberg, 2007).

\bibitem{us04}
U.\ Schwingenschl\"ogl and V.\ Eyert,
Ann.\ Phys.\ (Leipzig) {\bf 13}, 475 (2004).

\bibitem{us06}
U.\ Schwingenschl\"ogl and C.\ Schuster,
Chem.\ Phys.\ Lett.\ {\bf 432}, 245 (2006)

\bibitem{us07}
U.\ Schwingenschl\"ogl and C.\ Schuster,
Eur.\ Phys.\ J. B {\bf 55}, 43 (2007).

\bibitem{lb76}
Landolt-B\"ornstein, III/7, e760, d7716 (Springer, Heidelberg, 1976)

\bibitem{wien2k}
P.\ Blaha, K.\ Schwarz, G.\ Madsen, D.\ Kvasicka, and J.\ Luitz,
WIEN2k, An Augmented Plane Wave + Local Orbitals Program for
Calculating Crystal Properties (TU Wien, 2001).

\bibitem{us07c}
U.\ Schwingenschl\"ogl and C.\ Schuster,
Europhys.\ Lett.\ {\bf 77}, 37007 (2007).

\bibitem{us07d}
U.\ Schwingenschl\"ogl and C.\ Schuster,
Appl.\ Phys.\ Lett.\ {\bf 90}, 192502 (2007).

\bibitem{king-smith94}
R.D.\ King-Smith and D.\ Vanderbilt,
Phys.\ Rev.\ B {\bf 49}, 5828 (1994).

\bibitem{hardman94}
P.J.\ Hardman, G.N.\ Raikar, C.A.\ Muryn, G.\ van der Laan, P.L.\ Wincott,
G.\ Thornton, D.W.\ Bullett, and P.A.D.M.A.\ Dale,
Phys.\ Rev.\ B {\bf 49}, 7170 (1994).

\bibitem{gemming06}
S.\ Gemming and G.\ Seifert, Acta Mat.\ {\bf 54}, 4299 (2006).


\end{thebibliography}
\end{document}